\documentclass[fleqn,10pt]{wlscirep}
\usepackage[utf8]{inputenc}
\usepackage[T1]{fontenc}
\usepackage{algorithm}
\usepackage{algpseudocode}
\usepackage{textgreek}
\usepackage{amssymb}
\usepackage{tikz}
\newcommand{\hollowcircle}{\tikz[baseline=-0.75ex]\node[circle, draw, inner sep=2pt] {};} 
\newcommand{\filledcircle}{\tikz[baseline=-0.75ex]\node[circle, fill, inner sep=2pt] {};} 
\newcommand{\hollowhex}{\tikz[baseline=-0.75ex]\node[regular polygon, regular polygon sides=6, draw, inner sep=2pt, shape border rotate=30] {};} 
\newcommand{\filledhex}{\tikz[baseline=-0.75ex]\node[regular polygon, regular polygon sides=6, fill, inner sep=2pt, shape border rotate=30] {};} 
\newcommand{\hollowstar}{\tikz[baseline=-0.75ex]\node[star, star points=5, star point ratio=2.25, draw, inner sep=1.3pt] {};} 
\newcommand{\filledstar}{\tikz[baseline=-0.75ex]\node[star, star points=5, star point ratio=2.25, fill, inner sep=1.3pt] {};} 
\newcommand{\hollowdiamond}{\tikz[baseline=-0.75ex]\node[regular polygon, regular polygon sides=4, draw, inner sep=1.5pt, shape border rotate=45] {};} 
\newcommand{\filleddiamond}{\tikz[baseline=-0.75ex]\node[regular polygon, regular polygon sides=4, fill, inner sep=1.5pt, shape border rotate=45] {};} 
\newcommand{\hollowtriangle}{\scalebox{0.85}{$\triangle$}} 
\newcommand{\filledtriangle}{$\blacktriangle$} 
\usetikzlibrary{shapes.geometric}

\title{Academic collaborations and movements towards successful careers in physics}

\author[1,*]{Mingrong She}
\author[2,3,4]{Jan Bachmann}
\author[2,4]{Fariba Karimi}
\author[1]{Leto Peel}
\affil[1]{Department of Data Analytics and Digitalisation, School of Business and Economics, Maastricht University, Maastricht, The Netherlands}
\affil[2]{Complexity Science Hub, Vienna, Austria}
\affil[3]{Department of Network and Data Science, Central European University, Vienna, Austria}
\affil[4]{Graz University of Technology, 8010 Graz, Austria}

\affil[*]{m.she@maastrichtuniversity.nl}

\begin{abstract}
Collaboration networks evolve throughout academic careers, yet few studies systematically examine how these network dynamics relate to long-term career success and mobility. Analysing 35,708 physicists' careers spanning at least 15 years, we use time series clustering to identify ten distinct evolution patterns of network size and clustering coefficient across career years 5 to 15. We report three key results. First, authors who begin with loosely connected networks and progressively tighten their networks while expanding network size during mid-career achieve the highest PI attainment rates, publication output, and citation impact. Second, despite different starting points, network evolution patterns associated with better outcomes converge toward moderate clustering by career year 15, suggesting an optimal balance between core team cohesion and diverse external connections. Third, mobility is positively associated with these successful network evolution patterns and remains positively associated with scientific outcomes even after controlling for network evolution patterns.
\end{abstract}

\begin{document}

\flushbottom
\maketitle

%
\thispagestyle{empty}

\section{Introduction}
Academic careers are shaped by the dynamic interplay between collaboration networks~\cite{she2025gender, van2021collaboration} and international mobility~\cite{momeni2022many}, both of which are considered to fundamentally influence scientific success~\cite{burt2004structural, sugimoto2017scientists}. Understanding how these factors interact and co-evolve over time is crucial for explaining career trajectories and informing academic policy.

Collaboration networks fundamentally affect how researchers engage with the scientific community, influencing access to ideas, expertise, and opportunities~\cite{burt2004structural}. Researchers increasingly work in teams rather than individually~\cite{wuchty2007increasing}, and these networks can vary substantially in two key dimensions: size, which reflects the number of unique collaborators, and clustering, which reflects the degree to which collaborators are interconnected. These structural variations shape career development by influencing information flow across research communities, facilitating trust development within research teams, and affecting researcher visibility and reputation~\cite{she2025gender, van2021collaboration, zappalà2025genderdisparitiesdisseminationacquisition, dies2025forecastingfacultyplacementpatterns}. However, collaboration networks are not static. They evolve continuously as researchers gain experience, move between countries, form new ties, and enter different institutional and disciplinary contexts. Despite this dynamic nature, how network characteristics evolve over academic careers and how different evolutionary patterns relate to mobility and long-term success have not been systematically examined.

International mobility is considered a critical mechanism through which researchers can actively shape their collaboration networks and career outcomes. This view underpins prominent funding programs such as the Marie Skłodowska-Curie Actions, Fulbright Program, JSPS fellowships, and China Scholarship Council program. Moving across institutions or countries exposes researchers to new research environments, funding systems, and intellectual communities, fundamentally altering their opportunity structures.
Extensive empirical evidence supports this perspective: mobile researchers consistently outperform their non-mobile peers, achieving higher publication output~\cite{momeni2022many, azoulay2017mobility, cruz2010mobility, deville2014career, fernandez2016productivity, horta2010navel, tartari2020another, baruffaldi2012return, basu2013some, gibson2014scientific, hoisl2007study, jonkers2008chinese, teichler2017internationally, audretsch2016university} and greater citation impact~\cite{momeni2022many, gibson2014scientific, sugimoto2017scientists, franzoni2014movers, abramo2022mobility}, developing broader research portfolios~\cite{petersen2018multiscale}, building expanded collaboration networks~\cite{jonkers2013research, liu2022movers, wang2019collaboration, scellato2015migrant, chinchilla2021mena}, and gaining improved access to research resources~\cite{canibano2008measuring}. 

The relationship between mobility, network evolution, and scientific success is often interdependent and complex. Mobility influences collaboration network formation and transformation~\cite{jonkers2013research, liu2022movers, wang2019collaboration, paraskevopoulos2020dynamics, fontes2013impact, petersen2021propagation}. While moving to new institutions introduces researchers to new collaborators, it potentially weakens ties with previous collaborators. This affects both network size and clustering patterns. These mobility-induced network changes, in turn, can shape subsequent scientific success, positively or negatively, by altering access to resources, trusted information, and job opportunities~\cite{burt2004structural, coleman1988social}. High-achieving researchers may receive more attractive mobility opportunities and may be better positioned to form strategic collaborations. However, being too mobile may come at the cost of network support needed to secure tenure. This complex interdependence suggests that understanding career trajectories requires examining how mobility patterns and network evolution jointly shape career outcomes over time.

Despite growing interest in the role of collaboration networks and mobility in shaping academic careers, our understanding of how these factors evolve and jointly influence long-term success remains limited. 
Recent studies have examined how individual scientific impact changes over time~\cite{sinatra2016quantifying, thelwall2021career}, how mobility affects collaboration network dynamics~\cite{paraskevopoulos2020dynamics, fontes2013impact}, and how network structure relates to career outcomes~\cite{li2022untangling, xu2019exploring}. 
However, these studies either examine network evolution without systematically linking it to long-term outcomes, or analyse network structure at single time points without capturing evolutionary trajectories.
As a result, key questions remain unanswered: Do distinct patterns of network evolution exist across researchers? How do these patterns relate to long-term success? For instance, it remains unclear whether researchers benefit more from expanding their network or from maintaining a close-knit circle of collaborators, and whether optimal strategies differ across career stages.
Moreover, how mobility combines with network evolution patterns to shape career trajectories remains unexplored.

To address these gaps, we investigate how collaboration networks evolve over authors' careers and how these evolutionary patterns, together with mobility, relate to long-term scientific success. We adopt a dynamic perspective, tracing network evolution from career year 5 to 15 among 35,708 physicists---a period capturing the critical transition from early to mid-career stages. We focus on two key structural features: network size (the number of unique coauthors) and clustering coefficient (the degree to which coauthors have collaborated with each other). By constructing time series of these features for each author and applying time series clustering, we identify ten distinct network evolution patterns. Analysing these patterns reveals three key findings.
First, network evolution patterns are strongly associated with scientific outcomes. We examine four outcome measures: likelihood of becoming a principal investigator (PI), time to PI attainment, total publication output, and citation impact. Patterns characterised by network growth combined with increasing clustering demonstrate better performance across most outcomes. Notably, authors who begin with loose networks and progressively tighten their networks while expanding network size during mid-career achieve better scientific outcomes.
Second, mobility at different career stages is associated with distinct network evolution trajectories. Authors with initially tight collaborative relationships exhibit lower early-career mobility, while those with initially loose networks show higher early-career mobility. Mid-career mobility patterns also vary across network evolution types, with network-growing patterns generally associated with higher mid-career mobility.
Third, even after controlling for network evolution patterns, mobility remains associated with scientific outcomes, though the relationship varies by pattern type and career stage.

\section{Methods}
We analysed the American Physical Society (APS) dataset, which includes $678,916$ publications from 19 Physical Review journals published between 1893 and 2020. We used the same name-disambiguated data and principal investigator (PI) definition from She et al.~\cite{she2025gender}, where a PI is an author with at least three last-author publications, excluding publications with alphabetic author ordering. The dataset contains $868,506$ unique authors.

\subsection{Author Selection Criteria}
We selected authors based on specific criteria to ensure the dataset was representative of active, collaborative authors with sufficient career trajectories to analyse the long-term impact of collaboration network evolution and mobility. Specifically, we applied the following criteria:

\begin{enumerate}
    \item We selected authors with a career length of at least 15 years, providing an adequate timeframe to study the evolution of collaboration networks from early to mid-career stages. 
    \item Authors who became PIs within five years were excluded to focus the analysis on typical career trajectories rather than on exceptionally rapid transitions.
    \item Authors were included only if the number of coauthors was greater than one to ensure that they had participated in more than a minimal amount of collaborative activity and to allow us to compute their clustering coefficient.
    \item We included only authors with three or more publications to focus on authors who were sufficiently active. This is also the minimum publication count to attain PI status.
\end{enumerate}

After applying these criteria, the final dataset included $35,708$ focal authors.

\subsection{Scientific Outcomes}
We examined how changes in collaboration patterns from early to mid-career are associated with academic success. To capture different dimensions of career achievement, we evaluated four outcome measures:

\begin{itemize}
    \item \textbf{Likelihood of being a PI} reflects whether an author attained an independent leadership position, as defined in the Author Selection Criteria (at least three last-author publications).
    \item \textbf{Time to become a PI} is defined as the number of years between the start of an author's career (year of first publication) and the year they first met our criteria for PI status.
    \item \textbf{Total number of publications} represents overall research output and is a widely used indicator of productivity.
    \item \textbf{Citations per paper} are used to evaluate average impact. Citations incorporate only other APS publications and were calculated using the methodology described by Huang et al.~\cite{huang2020historical}, which counts citations accrued within 10 years of publication for each paper, excluding self-citations. To ensure comparability across different publication years, citation counts were normalised by dividing each paper’s citation number by the average citation count of papers published in the same year.
\end{itemize}

\subsection{Evolution of Collaboration Networks}
We analysed the evolution of collaboration networks using two primary statistics for each career year between year 5 and year 15:
\begin{itemize}
    \item Network Size: The cumulative count of unique collaborators an author engaged with until the respective career age.
    \item Clustering Coefficient: A statistic quantifying the interconnectedness among an author’s collaborators,
    defined as the ratio of collaborations between coauthors to all possible coauthor pairs.
\end{itemize}

To control for the influence of productivity on the collaboration networks, we regressed each network statistic on the number of publications separately for each career year from year 5 to year 15. This controls for the natural tendency of network size to increase and clustering coefficient to decrease with publication count.
For network size, we used a linear regression model since network size is continuous, and for clustering coefficient, we applied a binomial generalised linear model given that clustering coefficient values are bounded between 0 and 1. The standardised residuals from these models were used as adjusted network statistics, reflecting deviations from expected values given the author's publication output (see Appendix~\ref{Productivity_Control}
for detailed information). Based on these adjusted network statistics, we constructed two annual time series for each author, one for network size and one for clustering coefficient, covering the 11-year period.

To group authors according to how their collaboration networks changed over time, we applied Time Series K-Means~\cite{tavenard2020tslearn}, which groups authors based on the similarity of their network size and clustering coefficient trajectories over the 11-year observation period.

\begin{figure}[ht!]
    \centering
    \includegraphics[width=\textwidth]{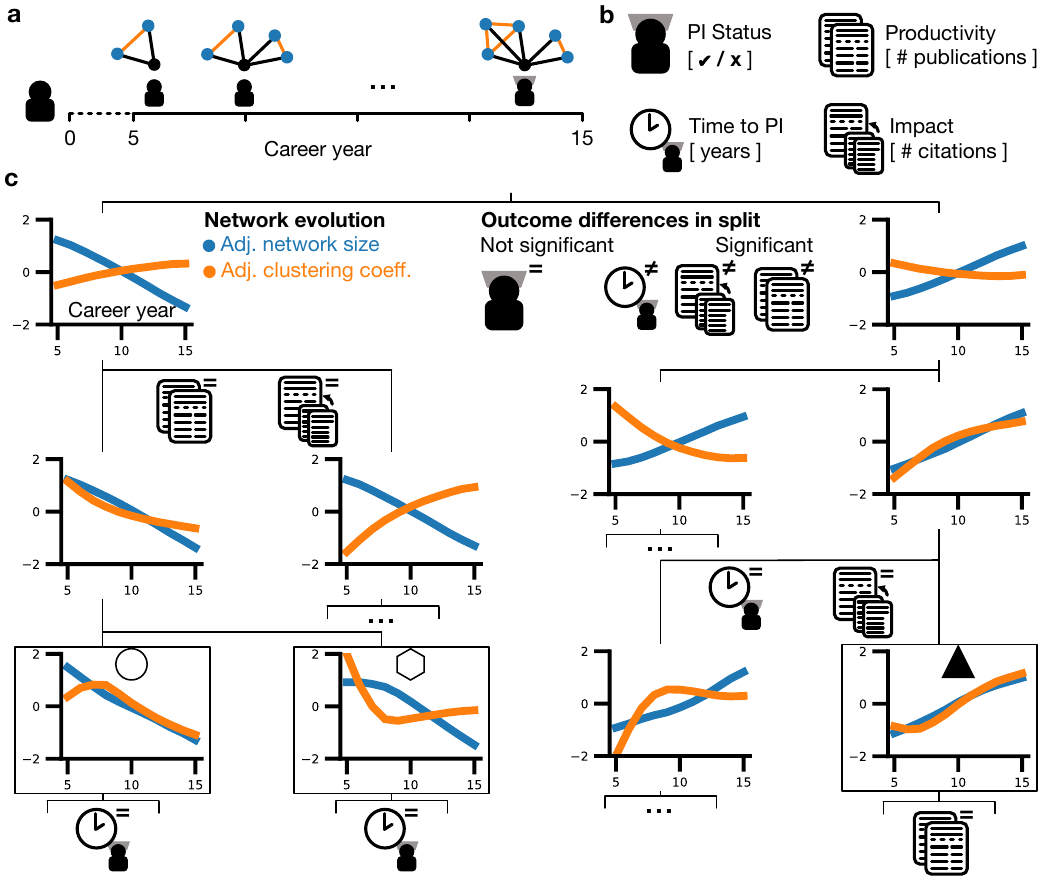}
    \caption{\textbf{Network evolution and outcome based clustering.} (A) We track authors' collaboration network evolution between the fifth and fifteenth year since their first publication. (B) Four outcome dimensions assess whether (i) scientists reached PI status, (ii) the time it took to reach it, (iii) their productivity and (iv) impact. (C) We apply Time Series K-Means clustering on the network evolution of all 35,708 authors to recursively assign them to one of two groups. Permutation tests detect significant differences between two groups in each of the four outcome dimensions. The splitting is repeated for each group until no significant difference can be detected for each of the four outcome variables. The last valid split is then considered as one of the final network evolution clusters (black border and marker).}
    \label{fig:figure_1}
\end{figure}

\subsubsection*{Recursive time series clustering}
Figure~\ref{fig:figure_1} illustrates our recursive clustering procedure. We began by applying Time Series K-Means with two clusters to the full sample, using both adjusted network size and adjusted clustering coefficient time series as input.
After the initial split, we compared the two resulting groups based on their scientific outcomes.
For each scientific outcome, we computed the observed difference in mean outcome values between the two groups and evaluated its significance using a Monte Carlo permutation test. We created 1,000 random partitions of authors into two groups of the same sizes as the identified clusters, computing the difference in means for each random partition.
The resulting p-value reflects the proportion of random splits with an equal or greater difference in means than the cluster assignment. If at least one outcome showed a statistically significant difference (p < 0.001), we recursively applied the same Time Series K-Means procedure to each subgroup. The recursive procedure follows a progressive filtering strategy: at the first level, all four outcomes are tested for significance; at each subsequent level, only outcomes that were significant in the immediate parent level are tested. This means that once an outcome becomes non-significant at any level, it is permanently excluded from all deeper recursive splits in that branch. The recursive process continued until no remaining outcome showed a statistically significant difference between the two subgroups, at which point the splitting of that branch was terminated. The recursive clustering algorithm is given in Appendix~\ref{clustering_algorithm}.
We obtained ten clusters using this process, each representing a distinct pattern of collaboration network development over time.

\begin{figure}[H]
  \centering
  \includegraphics[width=\textwidth]{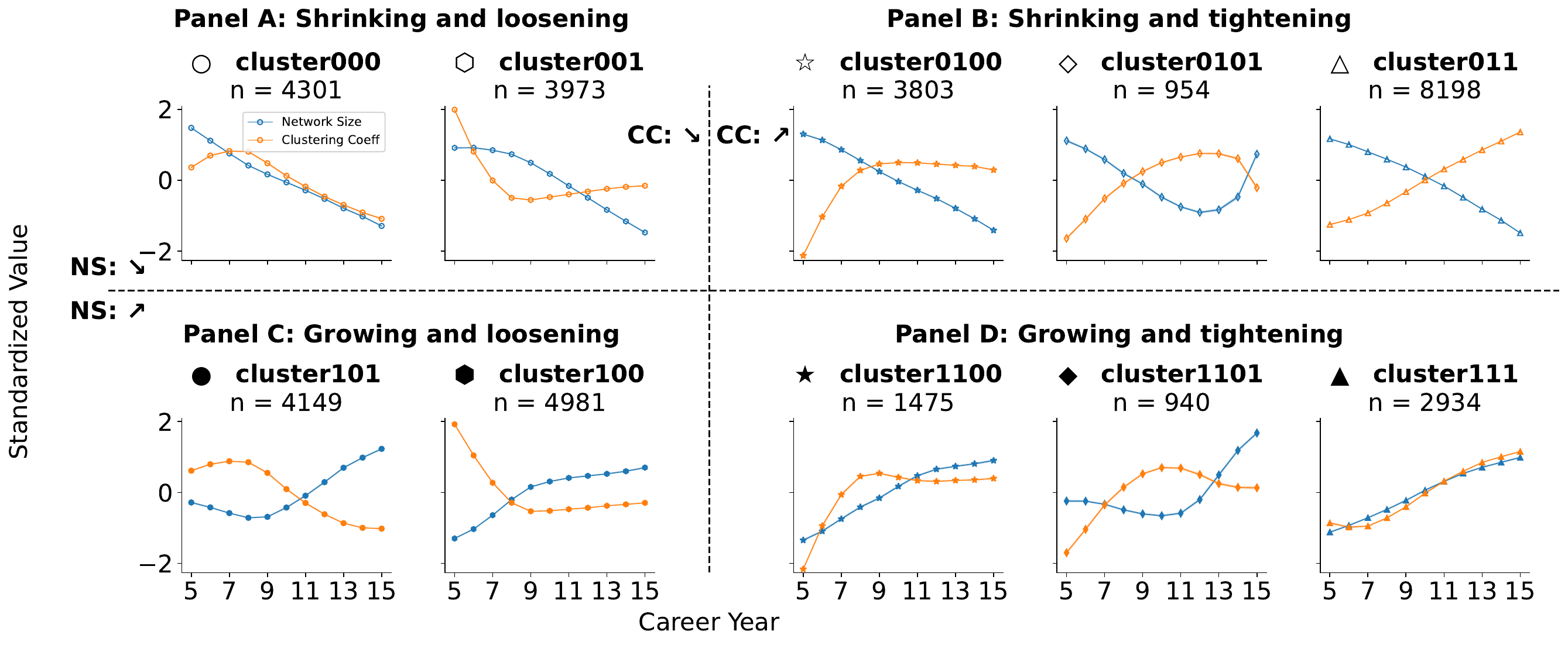}
    \caption{\textbf{Evolution of adjusted network size and clustering coefficient across 10 clusters.} Each subplot shows the trajectories of one cluster for adjusted network size (NS, blue solid line) and adjusted clustering coefficient (CC, orange solid line) over career years 5--15. Solid/hollow markers indicate adjusted network size trends (increasing/decreasing); marker shapes represent clustering coefficient patterns. The vertical dashed line separates clusters by clustering coefficient trends (left: $\searrow$ decreasing, right: $\nearrow$ increasing), while the horizontal dashed line separates clusters by network size trends (bottom:$\nearrow$ increasing, top: $\searrow$ decreasing). The figure is divided into four quadrants based on trajectory trends: Panel A (shrinking and loosening, top-left): NS$\searrow$, CC$\searrow$; Panel B (shrinking and tightening, top-right): NS$\searrow$, CC$\nearrow$; Panel C (growing and loosening, bottom-left): NS$\nearrow$, CC$\searrow$; Panel D (growing and tightening, bottom-right): NS$\nearrow$, CC$\nearrow$. }
  \label{fig:network_evolution_patterns}
\end{figure}

Figure~\ref{fig:network_evolution_patterns} shows the ten distinct network evolution patterns inferred by our algorithm, capturing diverse trajectories of adjusted network size and clustering coefficient over career years 5--15 (the absolute network values and productivity-based predictions are provided in Appendix~\ref{Actual_Network}).
The ten clusters are organised into four quadrants based on their network evolution trajectories. Clusters are labelled using binary notation reflecting the recursive splitting process: the first digit indicates adjusted network size trends (0 for decreasing, 1 for increasing), the second digit indicates clustering coefficient trends (0 for decreasing, 1 for increasing), and additional digits distinguish specific trajectory shapes.
Because these patterns are based on productivity-adjusted residuals rather than absolute values, the terms we use to describe them are relative: `shrinking' indicates network size growing slower than expected given publication output, while `growing' indicates faster-than-expected expansion. Similarly, `loosening' and `tightening' describe clustering coefficient changes relative to productivity-based expectations. We use this terminology throughout the remainder of the paper.
Markers encode network evolution patterns visually:
\begin{itemize}
    \item Hollow markers (top row) indicate \emph{shrinking} patterns; filled markers (bottom row) indicate \emph{growing} patterns
    \item \hollowcircle/\filledcircle~Circle: gradual \emph{loosening}, maintaining high clustering early before declining
    \item \hollowhex/\filledhex~Hexagon: rapid \emph{loosening}, quickly transitioning from tight to moderate-density networks
    \item \hollowstar/\filledstar~Star: rapid \emph{tightening}, quickly stabilising at high clustering
    \item \hollowdiamond/\filleddiamond~Diamond: moderate \emph{tightening}, peaking mid-career before declining somewhat
    \item \hollowtriangle/\filledtriangle~Triangle: continuous \emph{tightening}, increasing throughout the observation period.
\end{itemize}

We also assessed whether the network evolution patterns were associated with gender composition. We examined whether gender composition differed across the 10 network evolution clusters using a chi-squared test. To account for the potential impact of large sample size on statistical significance, we complemented the chi-squared test with Cram\'er's V as a measure of practical effect size. Results showed $\chi^2$ = 28.56 (df = 9, p = 0.0008) but V = 0.037, indicating that while differences were statistically detectable, the effect size was negligible (V < 0.1). Female representation ranged from 6.26\% to 9.30\% across clusters, confirming balanced gender composition across the network evolution patterns.

\section{Results}
\subsection{Network Evolution and Scientific Outcomes}
\begin{figure}[H]
    \centering
    \includegraphics[width=\textwidth]{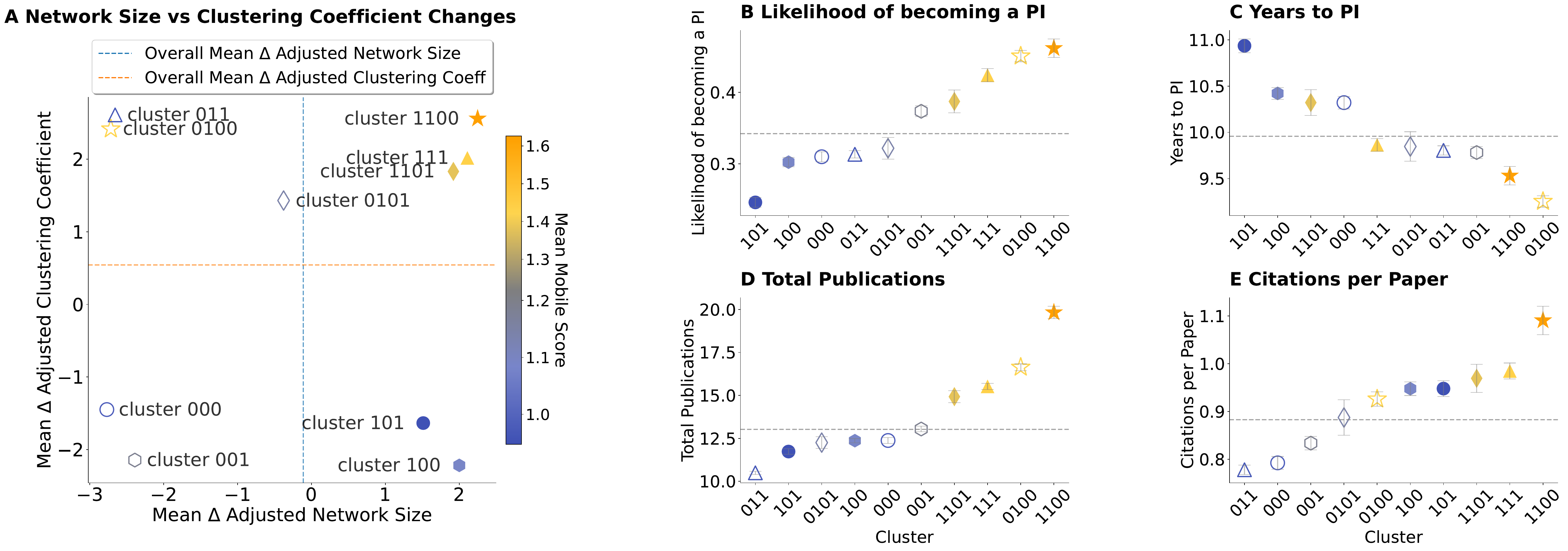}
    \caption{\textbf{Network evolution patterns and associated scientific outcomes.} Panel A shows mean changes in adjusted network size and adjusted clustering coefficient from career year 5 to year 15 for each cluster. Each point represents a distinct network evolution pattern, as shown in Figure~\ref{fig:network_evolution_patterns}, with reference lines showing overall means. Colours indicate mean mobility score for each cluster (indigo: low mobility, grey: medium mobility, gold: high mobility). Panels B-E compare scientific outcome measures across clusters: (B) likelihood of becoming a PI, (C) time to become a PI, (D) total publication output, and (E) citations per paper. Error bars represent standard errors of the mean. Clusters are ordered by performance within each outcome measure and the reference lines indicate the global mean.}
    \label{fig:cluster_outcomes}
\end{figure}

We examined how network evolution patterns relate to scientific outcomes using the four outcome measures described in the Methods section. To explore whether mobility is associated with these patterns, we also calculated the frequency of country transitions across each author's career using the
international mobility score described by Momeni et al.~\cite{momeni2022many}. 
For each publication year, we assigned one point for each new country not present in the previous year's affiliation list. For the first publishing year, we counted the number of unique countries and subtracted one to exclude the origin country.
Figure~\ref{fig:cluster_outcomes} shows the relationships between network evolution patterns, mobility, and scientific outcomes. Panel A shows the mean changes in adjusted network size and adjusted clustering coefficient from career year 5 to year 15 for each cluster, with colours indicating average mobility scores. Panels B-E compare the four scientific outcome measures across clusters.

\paragraph{Overall performance across quadrants.}
Overall, growing and tightening patterns (bottom-right quadrant: Clusters~\filledstar, \filleddiamond, \filledtriangle), characterised by both increasing clustering coefficient and increasing network size, had the highest mobility scores and achieved the best outcomes. These patterns all show network expansion combined with progressively increasing clustering. Among these, Cluster~\filledstar{} demonstrates the highest overall performance, ranking at the top in likelihood of becoming a PI, total publication output, and citations per paper. Clusters~\filleddiamond{} and \filledtriangle{} also perform well across most outcomes, though they are associated with longer time to achieve PI status.

\paragraph{Effects of network size and clustering.}
To understand the contribution of each network dimension, we examined network size and clustering coefficient evolution separately. Network size growth shows a consistent association with citation impact: regardless of clustering coefficient trajectories, patterns with growing adjusted network size trends (bottom quadrants) consistently show higher citations per paper across outcome comparisons. 
The association between clustering coefficient evolution and outcomes is more complex and appears to depend on network size trajectories. Among tightening patterns (right quadrants, where clustering coefficient increases over time), network size trajectory is associated with different outcome profiles. Growing and tightening patterns (bottom-right) compared to shrinking and tightening patterns (top-right) show higher PI attainment rates, greater total publication counts, and higher citations per paper, though they are associated with longer time to achieve PI status. This pattern is consistent across matched pairs within each tightening trajectory: \filledstar{} vs \hollowstar, \filleddiamond{} vs \hollowdiamond, and \filledtriangle{} vs \hollowtriangle.
Conversely, among loosening patterns (left quadrants, where clustering coefficient decreases over time), a different relationship emerges. Shrinking and loosening patterns (top-left) compared to growing and loosening patterns (bottom-left) show higher PI attainment rates, faster PI transition times, and greater publication output, though with lower citation impact per paper.

\paragraph{Outcomes among clusters with similar initial characteristics.}
Comparing clusters with similar initial network characteristics reveals how different evolutionary trajectories from comparable starting points relate to outcomes. Among authors in shrinking and loosening patterns (top-left), more rapid transition to moderate clustering while reducing network size is associated with better outcomes than gradual loosening. For authors in shrinking and tightening patterns (top-right), rapid development toward tight collaboration is associated with higher performance compared to gradual or moderate tightening. Among authors in growing and loosening patterns (bottom-left), network expansion combined with rapid transition to moderate clustering is associated with better outcomes than gradual loosening. Finally, for authors in growing and tightening patterns (bottom-right), rapid development from loose to tight collaboration while simultaneously expanding network size is associated with the highest performance compared to more gradual tightening trajectories. These results suggest that despite different starting points, patterns associated with better outcomes converge toward moderate clustering by career year 15.

\subsection{Network Evolution and Mobility}
The previous results showed that successful network evolution patterns tend to exhibit higher mobility scores. To better understand how mobility relates to network evolution at different career stages, we computed two specific mobility measures corresponding to the network evolution observation period. \textit{Early-career mobility} was calculated as the cumulative mobility score from career year 1 to year 5, capturing international transitions during the initial career phase before our network evolution observation period. \textit{Mid-career mobility} represented the cumulative mobility score from career year 6 to year 15, capturing transitions during the same period as our network evolution analysis. 

\begin{figure}[H]
    \centering
    \includegraphics[width=0.7\textwidth]{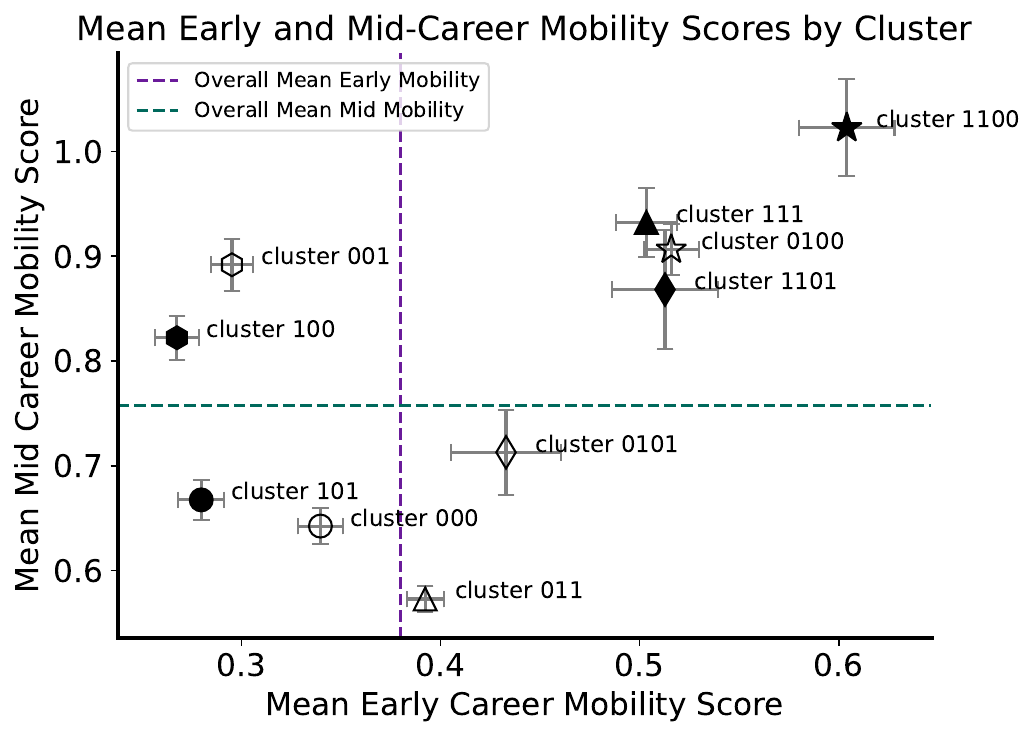}
    \caption{\textbf{Early and mid-career mobility patterns across network evolution clusters.} The scatter plot shows mean mobility scores with error bars indicating standard errors. Each point represents a cluster's average early-career mobility (x-axis) and mid-career mobility (y-axis). Reference lines indicate overall population means. Cluster markers correspond to network evolution patterns shown in Figure~\ref{fig:network_evolution_patterns}.}
    \label{fig:mobility_patterns}
\end{figure}

Figure~\ref{fig:mobility_patterns} presents the relationship between early-career and mid-career mobility scores across the ten network evolution clusters.
Our analysis revealed two key relationships between network evolution patterns and mobility.
First, early-career mobility shows a strong negative correlation with initial network density. Patterns in the left quadrants (shrinking and loosening, growing and loosening: Clusters~\hollowcircle, \hollowhex, \filledhex, \filledcircle), which have early tight collaborative relationships, demonstrate significantly lower early-career mobility scores. Conversely, patterns in the right quadrants (shrinking and tightening, growing and tightening: Clusters~\hollowstar, \hollowdiamond, \hollowtriangle, \filledstar, \filleddiamond, \filledtriangle), which have initially loose network structures, exhibit higher early-career mobility scores.
Second, mid-career mobility shows a general positive correlation with network size growth patterns, though with notable exceptions. Growing patterns (bottom quadrants), whether tightening or loosening, generally show higher mid-career mobility scores than shrinking patterns (top quadrants). Growing and tightening patterns (bottom-right: Clusters~\filledstar, \filleddiamond, \filledtriangle) consistently demonstrate both strong network size growth and higher mid-career mobility scores. Among growing and loosening patterns (bottom-left), Cluster~\filledhex{} shows higher mid-career mobility, while Cluster~\filledcircle{} represents an important exception, showing network size growth but relatively lower mid-career mobility. (Further details in Appendix~\ref{Mobility_network}).

\subsection{Mobility Associations with Scientific Outcomes within Network Evolution Patterns}
The previous sections showed that network evolution patterns and mobility are associated with scientific outcomes (as shown in Figure~\ref{fig:cluster_outcomes}) and that mobility varies within network evolution clusters (as shown in Figure~\ref{fig:mobility_patterns}). To investigate whether mobility remains associated with scientific outcomes after controlling for network evolution patterns, we conducted regression analyses within each cluster rather than comparing outcomes across different patterns. 
For binary outcomes (likelihood of being a PI), we applied logistic regression models within each cluster. For continuous outcomes (time to PI, total publications, citations per paper), we used ordinary least squares regression models.

\begin{figure}[H]
    \centering
    \includegraphics[width=\textwidth]{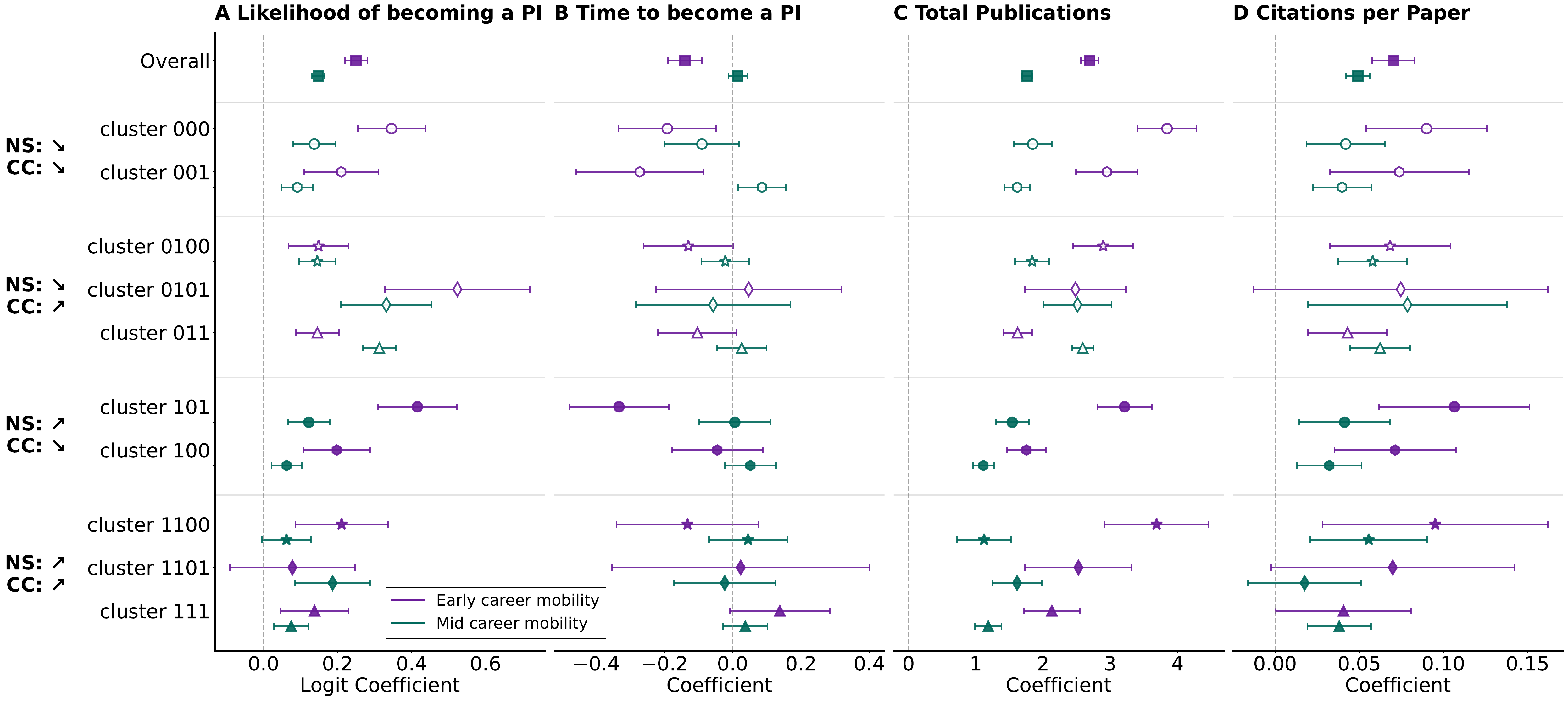}
    \caption{\textbf{Mobility effects on scientific outcomes across network evolution clusters.} Clusters are grouped by network evolution based on changes in adjusted network size (NS: $\nearrow$ increasing, $\searrow$ decreasing) and adjusted clustering coefficient (CC: $\nearrow$ increasing, $\searrow$ decreasing). Forest plots show regression coefficients with 95\% confidence intervals for early-career mobility (purple lines) and mid-career mobility (teal lines) effects within each cluster. Panel A: Logistic regression coefficients for likelihood of becoming a PI. Panel B: Linear regression coefficients for years to PI attainment. Panel C: Linear regression coefficients for total publications. Panel D: Linear regression coefficients for citations per paper.}
    \label{fig:mobility_forest_plots}
\end{figure}

Figure~\ref{fig:mobility_forest_plots} presents regression coefficients and 95\% confidence intervals for mobility associations with outcomes within each cluster. Overall, mobility is positively associated with scientific outcomes across most clusters. Both early-career and mid-career mobility show positive associations with likelihood of becoming a PI, total publications, and citations per paper in most clusters, while associations with time to PI are significant in only a few clusters.
The strength and timing of mobility associations vary across network evolution patterns. Clusters~\hollowcircle{} and \filledcircle, characterised by initially tight collaborative relationships transitioning to looser networks, show stronger associations between early-career mobility and both PI attainment and publication output compared to mid-career mobility. Conversely, Cluster~\hollowtriangle, characterised by a continuously tightening clustering coefficient, shows the opposite pattern, with mid-career mobility demonstrating stronger associations with both outcomes. Cluster~\hollowhex, which transitions from very tight to stable moderate-density networks, shows contrasting patterns for time to PI: early-career mobility is associated with shorter time, while mid-career mobility is associated with longer time.

\section{Discussion}
Here we examined how collaboration networks evolve over academic careers and their relationships with mobility and scientific success.
Our analysis shows that certain collaboration patterns are associated with better scientific outcomes. These successful patterns are characterised by authors who begin with loose collaborative networks in early career, then progressively strengthen these relationships while simultaneously expanding network size during mid-career. Additionally, despite starting with different network structures at career year 5, the patterns associated with better outcomes all move toward moderate clustering by career year 15. Authors starting with loose networks (low clustering coefficient) increase density during mid-career, while those starting with tight collaborative relationships (high clustering coefficient) decrease density. This convergence reveals that moderate density, which balances core team cohesion with diverse external connections, is key for better academic outcomes.
We also found that these successful patterns are associated with higher mobility at both early and mid-career stages. Even after controlling for network evolution patterns, mobility remains positively associated with scientific outcomes. 

The finding that patterns combining network expansion with increasing clustering are associated with better outcomes is consistent with existing theoretical perspectives on how networks may facilitate academic success. Network expansion provides access to diverse knowledge, resources, and opportunities~\cite{burt2004structural, granovetter1973strength}. At the same time, increasing network clustering may enable the development of stable collaborations necessary for sustained research projects~\cite{coleman1988social} and securing jobs through network support, though prior work suggests small and large teams serve different functions in scientific progress~\cite{wu2019large}. The convergence toward moderate rather than maximal clustering is particularly notable. Prior research suggests that extremely high clustering may restrict information flow~\cite{lindenlaub2021network}, while overly sparse networks may lack stable relationships for deep collaboration. The moderate clustering observed in high-performing patterns in our data appears consistent with a balance between these extremes. 

While prior research has documented general mobility benefits~\cite{momeni2022many, jonkers2013research, liu2022movers, wang2019collaboration}, our findings reveal that mobility is specifically associated with trajectories combining network expansion and increasing clustering. The nature of this association is unclear. Mobility may facilitate access to diverse collaborators that support these particular evolutionary patterns. Alternatively, mobility may be a marker of other characteristics associated with success. The relationship may also be bidirectional: successful network evolution may create mobility opportunities, while mobility may enable network development.

Our findings suggest potential considerations for researchers, mentors, and institutions. For individual researchers, our findings suggest that effective strategies may depend on career stage and current network configuration. Early-career researchers in our sample who developed better outcomes tended to build diverse collaborative networks. Authors with higher early-career mobility showed larger, more diverse networks. Mid-career appears to represent a transition point in our data, where successful patterns show both network expansion and increasing clustering. By later career, successful patterns maintain moderate clustering while expanding, suggesting the importance of balancing core collaborations with diverse connections. For mentors and advisors, our findings suggest that effective guidance may depend on career stage. Early-career mentees may benefit from encouragement to explore broadly~\cite{unger2022benefits}, while mid-career mentees may need help developing core collaborative teams while maintaining network growth. For institutions and funding agencies, our findings highlight the importance of supporting dynamic network evolution. Mobility programs appear associated with successful network development in our data. Long-term funding may enable the sustained trajectories needed for optimal network formation, avoiding premature specialisation caused by short-term contracts.

Several limitations of our paper point to directions for future research. First, our analysis relies on publication-based collaboration networks, which may not fully capture informal academic relationships such as mentorship or conference networking; future studies could broaden collaboration measurement to include research grants, mentorship ties, and conference interactions. Second, while the findings show strong associations between collaboration structures, mobility, and career success, causality remains uncertain due to potential selection effects; causal inference methods could clarify whether network structures drive success or reflect pre-existing heterogeneity. Third, our analysis focuses on career years 5 to 15, leaving earlier dynamics unexplored; future work could examine finer-grained trajectories and earlier career stages. Finally, our findings are based on physics and may not generalise to fields with different collaboration norms; cross-disciplinary comparisons could identify universal principles and field-specific dynamics. Despite these limitations, our work provides the first systematic examination of how collaboration network evolution and mobility jointly shape long-term academic careers. By moving beyond static snapshots of network structure, it opens new avenues for understanding the dynamic processes underlying scientific success and for developing evidence-based support for researchers throughout their careers.

\bibliography{references}

\section*{Acknowledgements}
MS acknowledges the financial support from the China Scholarship Council under grant number 202108080241. JB was supported by the Austrian Science Promotion Agency FFG under project No. 873927 ESSENCSE.
JB is a recipient of a DOC Fellowship of the Austrian Academy of Sciences at the Complexity Science Hub. LP was supported in part by the Dutch Research Council (NWO) Talent Programme ENW-Vidi 2021 under grant number VI.Vidi.213.163.

\appendix
\renewcommand{\thefigure}{\thesection.\arabic{figure}}
\renewcommand{\thetable}{\thesection.\arabic{table}}
\setcounter{figure}{0}
\setcounter{table}{0}

\section{Citation Measurement}
\label{citation_methodology}
To measure scientific impact consistently across different publication years and career stages, we applied the citation counting and normalisation methodology developed by Huang et al. (2020)~\cite{huang2020historical}. 
We removed all self-citations. A citation was identified as a self-citation when at least one author appeared on both the citing paper and the cited paper, based on author name disambiguation.
Citation counts were calculated using a fixed 10-year window. For each paper published in year $t$, we computed $c_{10}$ as the total number of non-self-citations received from year $t$ to year $t+10$. This fixed window addresses the issue that citations accumulate at different rates across papers~\cite{wang2013quantifying}.

To account for the fact that the average number of citations has changed over time~\cite{sinatra2016quantifying}, we normalised citation counts relative to the mean citation rate of papers published in the same year. For each paper $i$ published in year $t$, the normalised citation score was calculated as:

\begin{equation}
\text{normalised}\_c_{10}^{(i)} = \frac{c_{10}^{(i)}}{\overline{c_{10}}(t)} 
\label{eq:citation_normalization}
\end{equation}

where $c_{10}^{(i)}$ is the 10-year citation count for paper $i$, and $\overline{c_{10}}(t)$ is the mean 10-year citation count of all papers published in year $t$ in the APS dataset.
Citations per paper for each author were calculated as the mean of normalised $c_{10}$ scores across all papers by that author.

\section{Productivity Control Models}
\label{Productivity_Control}
To ensure that our clustering analysis captures network evolution patterns rather than productivity-driven effects, we controlled for the influence of publication output on network statistics through regression models applied separately for each career year. 

\begin{figure}[H] 
    \centering 
    \includegraphics[width=0.9\textwidth]{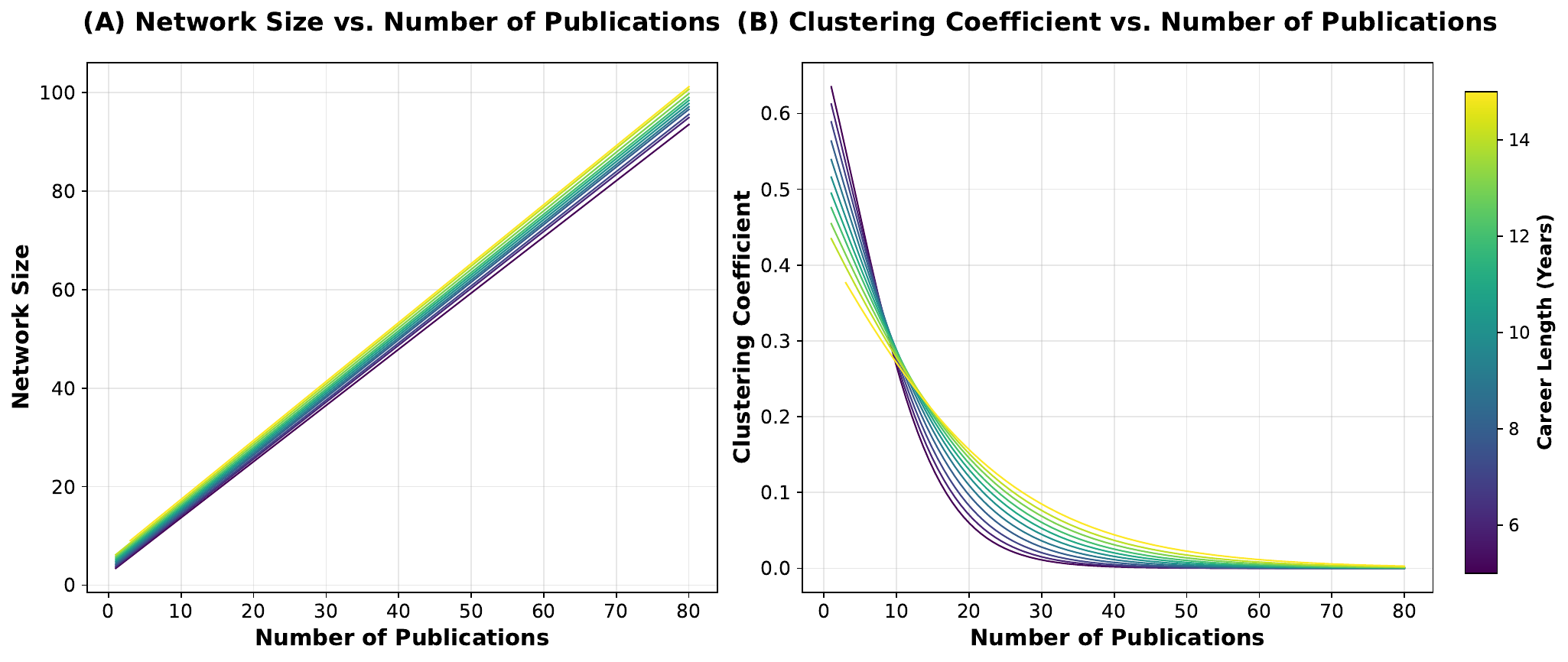} 
    \caption{\textbf{Regression relationships between network statistics and publication output across career stages.} Panel A: Network size versus number of publications fitted with ordinary least squares (OLS) models showing consistent linear relationships. Panel B: Clustering coefficient versus number of publications fitted with binomial generalised linear models (GLM) showing exponential decay patterns. Colours represent different career years from 5 (purple) to 15 (yellow) as indicated by the colour bar. These relationships were used to extract residuals for productivity-adjusted time series analysis.} 
    \label{fig:productivity_control} 
\end{figure}

Figure~\ref{fig:productivity_control} shows the regression results for our productivity control models. Panel A shows a consistent linear relationship between network size and number of publications across career years 5--15, with network size increasing approximately one collaborator per additional publication. Panel B shows the exponential decay relationship between clustering coefficient and publication output, where clustering coefficient approaches zero asymptotically as productivity increases. Early-career authors show slightly higher clustering coefficients at equivalent productivity levels, suggesting tighter collaborative relationships in initial career stages. These regression relationships were used to extract residuals that control for productivity effects in our time series clustering analysis. 

\section{Recursive Clustering Algorithm}
\label{clustering_algorithm}
We developed a recursive clustering approach to identify distinct network evolution patterns. The algorithm progressively partitions authors using Time Series K-Means, continuing to split subgroups only when they show statistically significant differences in scientific outcomes. This outcome-based validation ensures that identified clusters represent meaningful distinctions in career trajectories. The formal procedure is described in Algorithm~\ref{alg:recursive_clustering}.

\begin{algorithm}[H]
      \caption{Recursive Time Series Clustering}
      \label{alg:recursive_clustering}
      \begin{algorithmic}[1]
      \State \textbf{Input:} Time series data $X$, outcome variables $Y$,
  significance threshold $\alpha = 0.001$
      \State \textbf{Initialize:} Active outcomes $\mathcal{O} = \{$PI status,
  time to PI, publications, citations per paper$\}$
      \Function{RecursiveCluster}{$X$, $Y$, $\mathcal{O}$} \Comment{Recursive function}
          \State Apply Time Series K-Means with $k=2$ to $X$
          \State Split data and outcomes: $X_{G_0}, Y_{G_0}$ and $X_{G_1}, Y_{G_1}$
          \State Initialize $\mathcal{O}_{sig} = \emptyset$
          \For{each outcome $o \in \mathcal{O}$}
              \State Compute $\Delta_o = |\mu_{Y_{G_0}}(o) - \mu_{Y_{G_1}}(o)|$
              \State Perform Monte Carlo permutation test on pooled outcomes $Y_{G_0}(o) \cup Y_{G_1}(o)$ (1,000 iterations)
              \State Compute p-value $p_o$
              \If{$p_o < \alpha$}
                  \State Add $o$ to $\mathcal{O}_{sig}$
              \EndIf
          \EndFor
          \If{$\mathcal{O}_{sig} \neq \emptyset$}
              \State \Call{RecursiveCluster}{$X_{G_0}$, $Y_{G_0}$,
  $\mathcal{O}_{sig}$}
              \State \Call{RecursiveCluster}{$X_{G_1}$, $Y_{G_1}$,
  $\mathcal{O}_{sig}$}
          \Else
              \State Terminate and assign final cluster label
          \EndIf
      \EndFunction
      \State \Call{RecursiveCluster}{$X$, $Y$, $\mathcal{O}$}
      \end{algorithmic}
  \end{algorithm}

\section{Actual Network Statistic Trajectories}
\label{Actual_Network}
\begin{figure}[H]
    \centering
    \includegraphics[width=\textwidth]{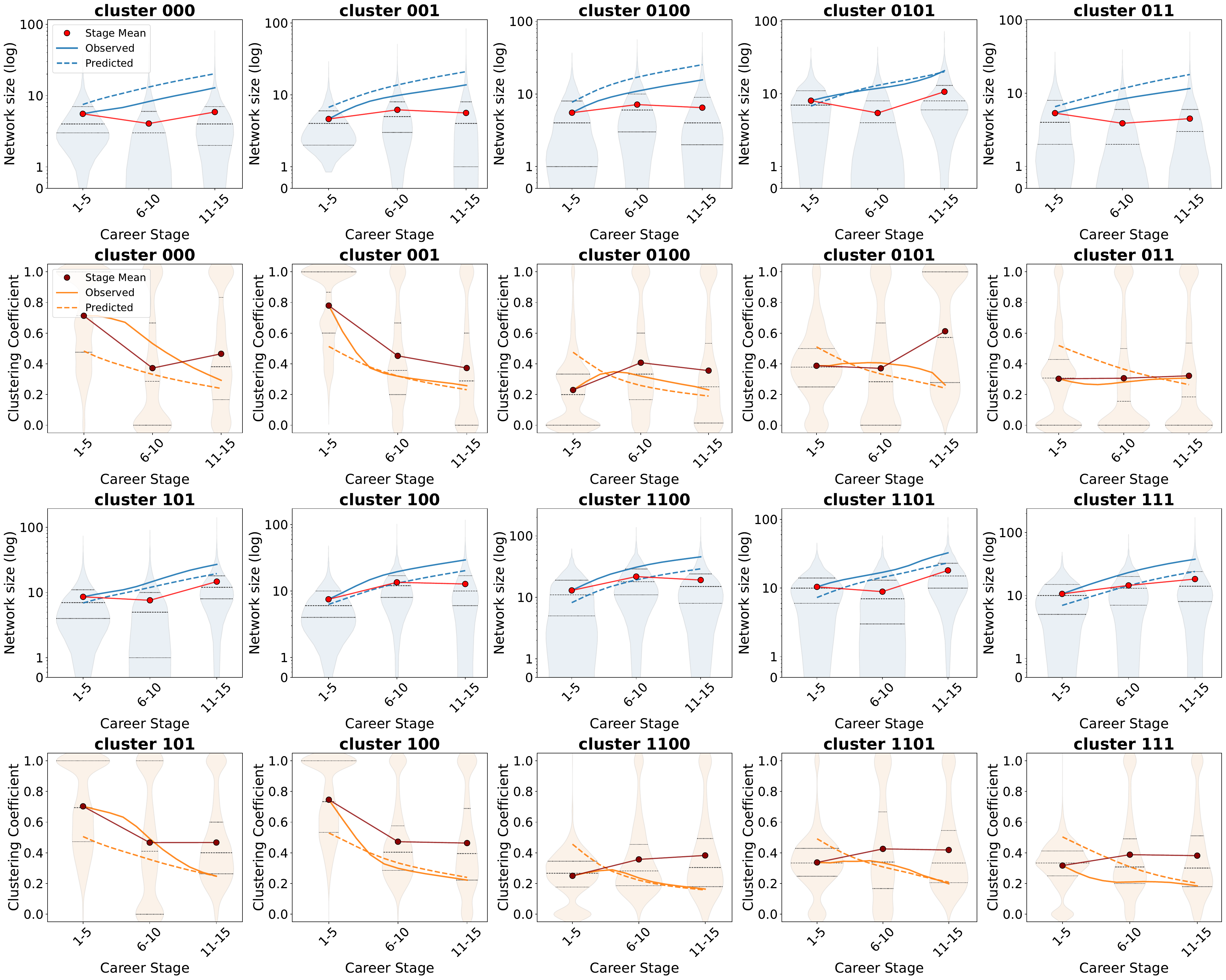}
    \caption{\textbf{Network evolution patterns across career trajectories and stages.} This figure shows the evolution of network size (rows 1 and 3) and clustering coefficient (rows 2 and 4) for 10 distinct career trajectory clusters across three career stages (1--5, 6--10, and 11--15 years). Violin plots show distributions, red markers indicate stage means, solid lines show observed values, and dashed lines show productivity-based predictions. The consistency between stage-wise and annual trends validates our temporal division.}
    \label{fig:stage_network_trajectories}
\end{figure}

The main analysis uses productivity-adjusted network statistics derived from regression models. Here we present the absolute values to show the actual network evolution patterns observed in our data. We analysed network size and clustering coefficient for authors in the ten career trajectory clusters across three career stages: early career (years 1--5), mid-career (years 6--10), and later career (years 11--15). Figure~\ref{fig:stage_network_trajectories} shows the distributions and mean values for each cluster at each stage, comparing observed values with productivity-based model predictions. The evolutionary patterns observed in these stage-wise means are consistent with the annual cumulative trends.

\section{Mobility and Network Structural Characteristics}
\label{Mobility_network}
To further examine the relationship between mobility patterns and network structural characteristics, we analysed how early-career mobility and mid-career mobility scores relate to adjusted network size and adjusted clustering coefficient across different mobility levels.

\begin{figure}[H]
    \centering
    \includegraphics[width=0.9\textwidth]{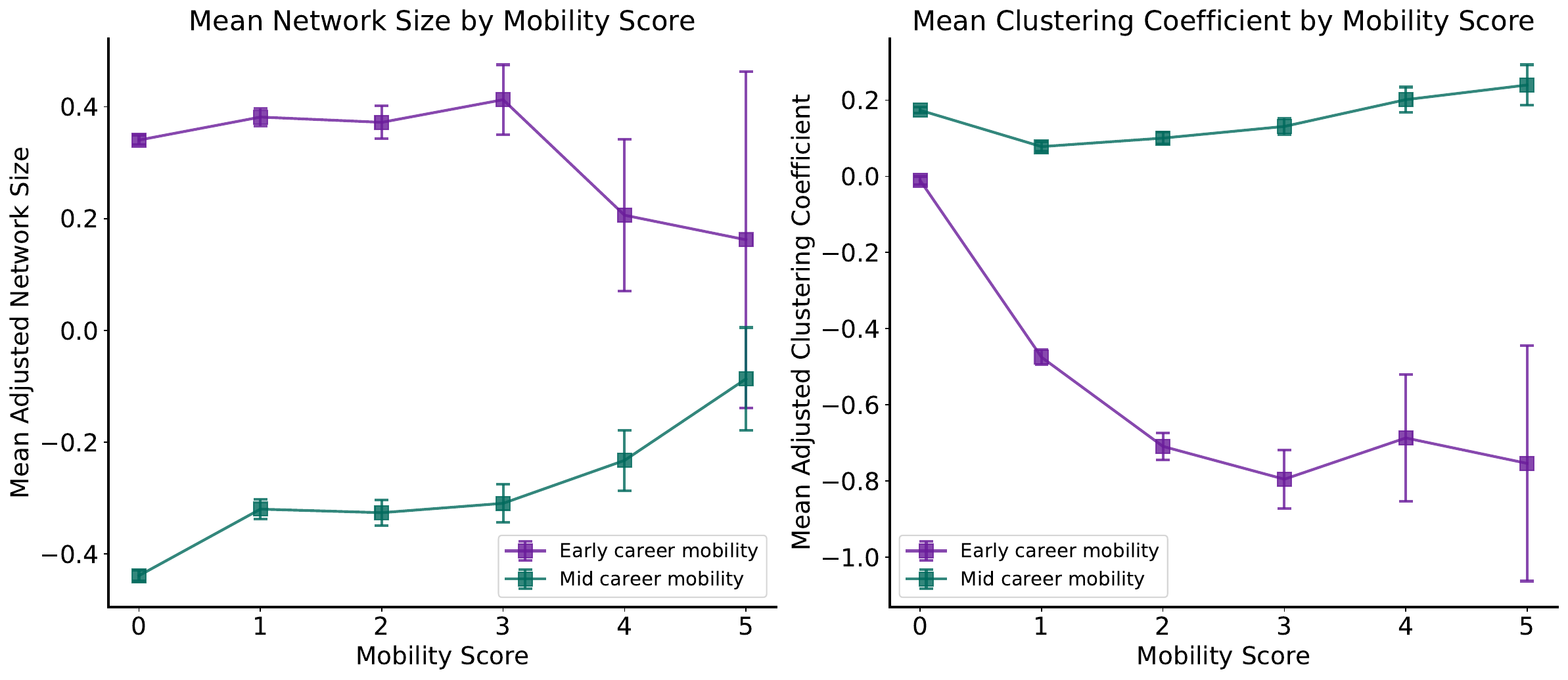}
    \caption{\textbf{Relationship between mobility scores and network statistics.} Left panel shows mean adjusted network size by mobility score for early-career mobility (purple circles) and mid-career mobility (teal squares). Right panel shows mean adjusted clustering coefficient by mobility score. Error bars represent standard errors of the mean. Only mobility scores with at least 10 observations are included.}
    \label{fig:network_clustering_mobility}
\end{figure}

Figure~\ref{fig:network_clustering_mobility} shows the mean adjusted network statistics grouped by mobility scores. For early-career mobility (measured during career years 1--5), adjusted network size remains relatively stable across mobility scores 0--3, then declines sharply at mobility scores 4--5. This indicates that authors with the highest early-career mobility scores have smaller network size. The adjusted clustering coefficient demonstrates a strong negative relationship with early-career mobility across all score levels, declining from approximately zero at score 0 to -0.8. This pattern indicates that authors with higher early-career mobility develop looser collaboration networks, where collaborators are less interconnected.

Mid-career mobility (measured during career years 6--15) exhibits different patterns. Adjusted network size shows a consistent positive relationship with mid-career mobility, increasing steadily from negative values at low mobility scores to positive values at high scores. This suggests that mid-career mobility facilitates network expansion beyond expected levels based on productivity. The adjusted clustering coefficient maintains relatively stable positive values across all mid-career mobility scores, ranging from approximately 0.05 to 0.20, indicating that mid-career mobility does not substantially affect the degree of interconnection among collaborators.

\end{document}